\documentclass[12pt]{iopart}
\begin{document}

\title[]{Post-Prior discrepancies in
CDW-EIS calculations for ion impact ionization fully differential
cross sections}

\author{M F Ciappina and W R Cravero}
\address{CONICET and Departamento de F\'{\i}sica, Av. Alem 1253, (8000) Bah\'{\i}a Blanca, Argentina}
\ead{ciappina@uns.edu.ar}

\begin{abstract}

\noindent In this work we present fully differential cross
sections (FDCSs) calculations using post and prior version of
CDW--EIS theory for helium single ionization by 100 MeV C$^{6+}$ amu$^{-1}$
and 3.6 MeV amu$^{-1}$ Au$^{24+}$ and Au$^{53+}$ ions. We performed our
calculations for different momentum transfer and ejected electron
energies. The influence of internuclear potential on the ejected
electron spectra is taken into account in all cases. We compare
our calculations with absolute experimental measurements. It is
shown that prior version calculations give better agreement with
experiments in almost all studied cases.

\end{abstract}

%Uncomment for PACS numbers title message
%\pacs{00.00, 20.00, 42.10}
% Keywords required only for MST, PB, PMB, PM, JOA, JOB?
%\vspace{2pc}
%\noindent{\it Keywords}: Article preparation, IOP journals
% Uncomment for Submitted to journal title message
\submitto{\JPB}
% Comment out if separate title page not required
\maketitle

\section{Introduction}

The study of electron emission spectra in ion atom collisions has
been a field of intense activity for years (Stolterfoht {\it et al} 1997). For
intermediate to high energy single ionization there has been
considerable theoretical efforts focused in the so-called two
centre electron emission (TCEE) (Fainstein {\it et al} 1991, Pedersen {\it et al} 1990).
Improvement in the description of the ionized electron moving in
the presence of both residual target and projectile fields after
the collision (final state) has been key for the correct
description of experimental data (Guly\'as {\it et al} 1995).

Within distorted wave approximations, it has been shown that, at
least for high impact energy and multiply charged projectiles, CDW
theory of Belki\'c (1978) used together with an appropriate
description of the initial bound and final continuum electron
states, yields best results for doubly differential cross sections
(DDCSs) (Guly\'as and Fainstein 1998, Ciappina {\it et al} 2003).
However, when the projectile impact velocity decreases, the
CDW--EIS theory of Crothers \& McCann (1983) gives better results,
its only difference being the choice of the initial state.
Moreover CDW--EIS approximation is formally free of criticisms
regarding the initial state proper normalization, and the
transition amplitudes have not the divergent behavior that CDW
exhibits (Crothers 1982) (although it has been demonstrated that
CDW amplitudes are integrable and its DDCSs are well behaved)
(Dub\'e and Dewangan 1995).

The field has experienced a renewed interest as a result of the
development of the experimental technique known as COLTRIMS
(cold-target recoil-ion momentum spectroscopy) (Moshammer {\it et
al} 1994). With COLTRIMS, the projectile's tiny scattering angle
can be obtained indirectly by measuring the ionized electron and
recoil ion momenta. Fully differential cross sections (FDCS) for
ion impact ionization can be measured now and this constitute a
challenging ground for existing theories (Foster {\it et al}
2004).

The first measurements of the FDCS, for various momentum transfers
and ejected-electron energies, were reported in 2001 by Schulz
{\it et al} for single ionization of helium by 100 MeV amu$^{-1}$
C$^{6+}$. Theoretical results for this process were made later by
Madison {\it et al} 2002, using several approximation schemes.
They obtained reasonable good agreement between experiment and
theory in the scattering plane for intermediate momentum transfer,
but the theories used was not able to reproduce the measurements
for large values of momentum transfer and out of scattering plane.

Subsequently, experiments with other projectiles and energy ranges
have been performed. Fischer {\it et al} (2003) have reported
absolute experimental measurements for 2 MeV amu$^{-1}$ C$^{6+}$
single ionization of helium in the scattering plane for various
momentum transfers and ejected-electron energies. Foster {\it et
al} (2004) have presented 3DW--EIS results for the single
ionization of helium by 3.6 MeV amu$^{-1}$ Au$^{24+}$ and
Au$^{53+}$ ions. The 3DW--EIS model is a modified version of the
CDW--EIS approximation and, although the authors found good
agreement with 2 MeV amu$^{-1}$ C$^{6+}$ data, the theory did not
yield a significant improvement for higher charged ions.
Theoretical results calculated using a CDW--EIS model exhibited
differences between experiment and theory on an absolute scale for
emission in the scattering plane, defined by the plane containing
the initial and final projectile momenta. Their calculations were
made using a post version of the CDW--EIS theory and an active
electron approximation with hydrogenic wavefunctions for the
initial and final electron states. Indeed, the simplest
description for the He bound initial state is to assume it has one
`active' and one `passive' electron and that the `active' electron
can be described as moving in the effective Coulomb field of the
atomic core with an effective charge chosen either: (a) to
reproduce the ionization energy or (b) so that the continuum wave
is orthogonal to the initial state.

A more sophisticated description involves the use of full
numerical Hartree--Fock wave functions for both initial and final
states of the active electron (Guly\'as {\it et al} 1995, Guly\'as
and Fainstein 1998, Foster {\it et al} 2004). Hartree--Fock
description, however, does not include proper angular correlations
in the initial state, and for large perturbations, there might be
the chance that the projectile interacts with more than one
electron in a single event. An explicit two-electron description,
i.e., a full-blown four-body theory for the collision process
might be necessary in that case. We have shown that by using the
prior version of CDW--EIS together with an appropriate
Roothan--Hartree--Fock description of the initial state and an
effective charge coulomb wave function for the target - electron
continuum, we get similar results to those obtained by using
numerical Hartree--Fock wave functions (Ciappina {\it et al}
2004), for ion impact helium ionization DDCSs.

The aim of this paper is to present post and prior CDW--EIS
calculations with internuclear interaction between the projectile
and the target (N--N interaction) taken into account for ion
helium single ionization FDCSs at different perturbation regimes.
Atomic units are used throughout unless otherwise stated.

\section{Theories}

We regard He single ionization as a single electron process and
assume that (i) the initial state for the `active' electron is
described by a semi-analytical Rothan--Hartree--Fock scheme using
a 5 parameters wave function (Clemente and Roetti 1974) and (ii)
in the final state the `active' target electron moves in the
combined Coulomb field of the target core with an effective charge
$Z_{eff}=1.6875$. The electron-projectile relative motion are
represented in a CDW--EIS approach, i.e. one eikonal phase in the
entrance channel and a pure Coulomb distortion in the final one.
N--N interaction is treated as a pure Coulomb interaction between
the projectile with a charge $Z_{P}$ and the true target core
charge, $Z_{T}$ = 1.

N--N interaction is taken into account in the transition amplitude
$a_{if}(\mbox{\boldmath$\rho$})$, in the usual semi-classical or
eikonal approximation, through its multiplication by a phase
factor (McCarroll and Salin 1978), which for pure coulomb
internuclear interaction results in (Crothers and McCann 1983)
\begin{equation}
a_{if}^{\prime}(\mbox{\boldmath$\rho$})=\mathrm{i}(\rho
v)^{2\mathrm{i}\nu}a_{if}(\mbox{\boldmath$\rho$})
\end{equation}
were $\nu=Z_{P}Z_{T}/v$, $v$ is the velocity of the impinging
projectile and $\rho$ is the impact parameter
$(\mbox{\boldmath$\rho$}\cdot\mathbf{v}=0)$. $ a_{if}\left(
\mbox{\boldmath$\rho$}\right)\;\left(
a_{if}^{\prime}\left(\mbox{\boldmath$\rho$}\right)\right)$ is the
transition amplitude with (without) internuclear interaction.
Using two-dimensional Fourier transforms we have for the
transition amplitude elements,CDW--EIS transition matrix can be
written alternatively as a function of the momentum transfer:

\begin{equation}
T_{if}^{\prime}(\mbox{\boldmath$\eta$}^{\prime})=\frac{\rmi
v^{2\rmi \nu}}{\left(2\pi\right)^{2}} \int{
d\mbox{\boldmath$\eta$}\ T_{if}(\mbox{\boldmath$\eta$})} \int{
d\mbox{\boldmath$\rho$}\; \rho^{2\mathrm{i}\nu} \
e^{\mathrm{i}(\mbox{\boldmath$\scriptstyle{\eta-\eta^{\prime}}$})
\cdot\mbox{\boldmath$\scriptstyle{\rho}$}} \
a_{if}(\mbox{\boldmath$\rho$})}
\end{equation}
We solve the integral over impact parameter analytically to
obtain:
\begin{equation}
T_{if}^{\prime}(\mbox{\boldmath$\eta$})=\nu\frac{\mathrm{i}v^{2\mathrm{i}\nu}
(2\pi)^{-\mathrm{i}\nu}}{2^{4}\pi^{3}} \int
d\mbox{\boldmath$\eta$}^{\prime}\ T_{if}
(\mbox{\boldmath$\eta$}^{\prime}) \left\vert
\mbox{\boldmath$\eta-\eta$}^{\prime}\right\vert
^{-2(1+\mathrm{i}\nu)} \label{Tif}%
\end{equation}

The remaining integral in (\ref{Tif}) is evaluated numerically
with an adaptive integration scheme. This approximation is valid
as long as (i) the projectile suffers very small deflections in
the collision and (ii) the velocity of the recoil ion remains
small compared to that of the emitted electron.

Within CDW--EIS, Transition amplitude can be computed as

\begin{equation}
T_{if}^{+CDW-EIS}=\left\langle \chi_{f}^{-CDW} \right|
W^{\dagger}_{f} \left| \chi_{i}^{+EIS}\right\rangle \label{post}
\end{equation}
in its post version or
\begin{equation}
T_{if}^{-CDW-EIS}=\left\langle \chi_{f}^{-CDW} \right| W_{i}
\left| \chi_{i}^{+EIS}\right\rangle \label{prior}
\end{equation}
in the prior version, where the initial (final) state distorted
wave $\chi_{i}^{+}$ ($\chi_{f}^{-}$) is an approximation to the
initial (final) state which satisfies outgoing-wave ($+$)
(incoming-wave ($-$)) conditions. For the initial state the
asymptotic form of the Coulomb distortion (eikonal phase) is used
in the electron-projectile interaction together with a semi
analytical Rothan--Hartree--Fock description for the initial
bound-state wavefunction (Clementi and Roetti 1974)
\begin{equation}
\chi_{i}^{+EIS}=(2\pi)^{-3/2}\exp\left(\mathrm{i}\mathbf{K}_{i}\cdot
\mathbf{R}_{T} \right)
\psi_{1s}(\mathbf{r}_{T})\mathcal{E}^{+}_{v}(\mathbf{r}_P)
\end{equation}
where $\mathcal{E}^{+}_{v}(\mathbf{r}_P)$ is
\begin{equation}
\mathcal{E}^{+}_{v}(\mathbf{r}_P)=\exp\left(-\mathrm{i}\frac{Z_{P}}{v}\ln
\left(v r_{P}- \mathbf{v}\cdot\mathbf{r}_{P} \right)\right).
\end{equation}

The final state wavefunction is collected into the form (Rosenberg
1973, Garibotti and Miraglia 1980, Crothers and McCann 1983)
\begin{equation}
\chi_{f}^{-CDW}=(2\pi)^{-3/2}\exp\left(\mathrm{i}\mathbf{K}_{f}\cdot
\mathbf{R}_{T} \right)
\chi_{T}^{-}(\mathbf{r}_{T})C_{P}^{-}(\mathbf{r}_{P})
\end{equation}
where $C_{P}^{-}$ represents the Coulomb distortion of the ejected
electron wave function due to the projectile coulomb potential.
\begin{equation}
C_{P}^{-}(\mathbf{r}_{P})=N(\nu_{P})
\left._1F_1\right.\left(-\rmi\nu_{P},1,-\rmi
k_{P}r_{P}-\rmi\mathbf{k}_{P}\cdot\mathbf{r}_{P}\right)
\end{equation}
being $\nu_{P}=\frac{Z_{P}}{k_{P}}$ the Sommerfeld parameter, and
$\mathbf{k}_{P}$ is the relative momentum of the e-P subsystem.
The $\left._1F_1\right.$ is the Kummer function and $N(\nu_{P})$
is the usual normalization factor
\begin{equation}
N(\nu_{P})=\Gamma(1-\rmi\nu_P)\exp(-\pi\nu_{P}/2)
\end{equation}
being $\Gamma$ the gamma function. On the other hand
$\chi_{T}^{-}(\mathbf{r}_{T})$ is the wave function for the
ejected electron in the field of the target residual ion.
\begin{equation}
\chi_{T}^{-}(\mathbf{r}_{T})=(2\pi)^{-3/2}\exp\left(\rmi\mathbf{k}_{T}\cdot\mathbf{r}_{T}\right)N(\nu_{T})
\left._1F_1\right.\left(-\rmi\nu_{T},1,-\rmi
k_{T}r_{T}-\rmi\mathbf{k}_{T}\cdot\mathbf{r}_{T}\right)
\end{equation}
being $\nu_{T}=\frac{Z_{T}}{k_{T}}$ and now $\mathbf{k}_{T}$ is
the relative momentum of the e-T subsystem. We use
$Z_{T}=Z_{eff}=1.6875$ to model the screened target residual ion
 as a pure Coulomb potential.

The perturbation potentials $W_{f}$ in equation (\ref{post}) and
$W_{i}$ in (\ref{prior}) are defined by
\begin{equation}
\left(H_{f}-E_{f}\right)\chi_{f}^{-}=W_{f}\chi_{f}^{-}
\end{equation}
and
\begin{equation}
\left(H_{i}-E_{i}\right)\chi_{i}^{+}=W_{i}\chi_{i}^{+}
\end{equation}
where $H_{f}$ ($H_{i}$) are the full electronic final (initial)
Hamiltonian (neglecting the total center of mass motion) and
$E_{f}$ ($E_{i}$) are the total final (initial) energy of the
system in the cm frame respectively.

The explicit forms of these operators can be written (Crothers and
Dub\'e 1992)
\begin{equation}
W_{f}=-\mbox{\boldmath$\nabla$}_{\mathbf{r}_{T}}\cdot\mbox{\boldmath$\nabla$}_{\mathbf{r}_{P}}
\end{equation}
and
\begin{equation}
W_{i}=\frac{1}{2}\mbox{\boldmath$\nabla$}^{2}_{\mathbf{r}_{P}}-
\mbox{\boldmath$\nabla$}_{\mathbf{r}_{T}}\cdot\mbox{\boldmath$\nabla$}_{\mathbf{r}_{P}}
\end{equation}

In the centre of mass frame, the FDCS in energy and ejection angle of the electron, and direction of the
outgoing projectile is given by (Berakdar {\it et al} 1993, Inokuti 1971, Bethe 1930)

\begin{equation}
\frac{\rmd^{3}\sigma}{\rmd E_{k}\rmd\Omega_{k}\rmd\Omega_{K}}=N_{e}(2\pi)^{4}\mu^{2}k\frac{K_{f}%
}{K_{i}}\left\vert T_{if}\right\vert ^{2}\delta(E_{f}-E_{i})
\label{FDCS}
\end{equation}

\noindent where $N_{e}$ is the number of electrons in the atomic
shell, $\mu$ is the reduced mass of the projectile-target
subsystem, $K_{i}\left( K_{f}\right)$ is the magnitude of the
incident particle initial (final) momentum. The ejected-electron's
energy and momentum are given by $E_{k}$ and $k$ respectively. The
solid angles $\rmd\Omega_{K}$ and $\rmd\Omega_{k}$ represent the
direction of scattering of the projectile and the ionized
electron, respectively. We use non-orthogonal Jacobi coordinates
$(\mathbf{r}_{P},\mathbf{r}_{T})$ to outline the collision
process. These coordinates are the position of the active electron
with respect to the projectile ($\mathbf{r}_{P}$) and to the
target ion ($\mathbf{r}_{T}$) respectively. Also the coordinate
$\mathbf{R}_{T}$ is needed, that represents the position of the
incoming projectile with respect to the center of mass of the
subsystem e-T. If we neglect terms of order $1/M_T$ and $1/M_{P}$,
where $M_{T}$ is the mass of the target ion nucleus and $M_{P}$ is
the corresponding to the incident heavy ion, we can write
$\mathbf{R}_{T}=\mathbf{r}_{T}-\mathbf{r}_{P}$.

We have replaced the transition matrix in the post and prior
schemes (equations (\ref{post}) and (\ref{prior})) in the
definition of FDCS (\ref{FDCS}) and we have applied it to several
single ionization processes.

\section{Results}

We have performed calculations for different projectiles, spanning
a large range of perturbation strengths as measured by charge to
velocity ratio $\eta=Z_{P}/v$. In figure 1 we present results for
100 Mev amu$^{-1}$ C$^{6+}$ (Schulz {\it et al} 2001) single
ionization of Helium calculated in prior CDW--EIS, for different
values of electron emission energy ($E_{e}$) and momentum transfer
($\mathbf{q}=\mathbf{K}_{i}-\mathbf{K}_{f}$). Calculations are in
very good agreement with available experimental results.

In figure 2 we layout results for other theories applied to the same process
for an intermediate value of electron energy and momentum
transfer. We see that prior CDW--EIS gives the best results. Even
when $\eta=0.1$ both FBA and post CDW--EIS fail to
accurately describe the experimental results, although they
broadly reproduce the angular distribution.

Figures 3 and 4 show results for 3.6 MeV amu$^{-1}$ Au$^{24+}$
impact ionization of He (Fischer {\it et al} 2003), calculated in
prior and post CDW--EIS. For $E_{e}=4.0$ eV, results
for prior version are in reasonable agreement with experiment,
However both theories fail to correctly reproduce the strong
forward emission peak, which is due to the strong projectile
electron post collisional interaction (PCI). Trend is similar for
$E_{e}=10.0$ eV, (Figure 4) where we see a better performance in
prior version calculations, in particular in the prediction of the
direct peak position. Note that no renormalization factor is
included in these calculations.

In figures 5 and 6 we show prior and post CDW--EIS calculations
for 3.6 MeV amu$^{-1}$ Au$^{53+}$ impact ionization of He. Even
when we are stretching the validity range of the perturbative
treatment a bit too much ($\eta \approx2,4.4$ for Au$^{24+}$ and
Au$^{53+}$ projectiles respectively), angular structure with only
one strong peak is correctly predicted in prior version while post
version of the theory predicts two distinct direct and recoil
peak. However the position of the peak is not correctly given in
prior version, again because the theory underestimates the strong
PCI between the impinging ion and the ejected electron, which
shifts the emission towards the forward direction. Both versions
including N-N interaction fail to yield the correct order of
magnitude of experimental data. Large projectile charges are
likely to induce quite a large polarization in the target.
Effective charges both for residual target-electron and N--N
interactions are probably not the same than for lower charged
projectiles, and it is indeed very probable that the effective
charge approach is not a good approximation here.  Model
potentials taking into account polarization effects need to be
considered for the target, at least in the exit channel, but most
probably in both initial and final states.

\section{Conclusions}

We have performed FDCSs calculations for highly charged ion impact ionization
of Helium. We employed prior and post versions of CDW--EIS theories taken into
account N--N interaction but otherwise using as simple an approach for
electronic wave functions as possible. Indeed, use of prior version helps us
to avoid the need of more precise wave functions for the initial or final
electronic state. We found reasonably good agreement with experimental data,
even for projectile charges for which the system is arguably outside the range
of validity of a perturbative theory.

We see that for emission in the collision plane, three body
dynamics seems to be enough to explain most of the structures
observed for low energy emission and low projectile charge. For
Au$^{24+}$ and Au$^{53+}$ projectiles the larger emission in the
forward direction is not well reproduced by the theory but, as
said before, those cases are outside the range where perturbative
treatments are known to be valid. However, if the effect of target
polarization in the entrance channel and the inclusion of higher
orders in the exit channel distortions, are taken into account,
perturbation based calculations could probably be brought closer
to experimental results.

\section{Acknowledgments}

This work has been partially supported by Consejo Nacional de Investigacones
Cient\'{\i}ficas y T\'{e}cnicas, Argentina, ANPCYT, PICT and Universidad
Nacional del Sur under PGI 24/F027. One of us (MFC) is grateful for the hospitality of the
Max Planck Institut f\"ur Kernphysik in Heidelberg.

\bigskip

\section*{References}

\begin{harvard}

\item[] Bethe H 1930 {\it Ann. Phys., Lpz} {\bf 5} 325
\item[] Belki\'c D\v{z} 1978 \JPB {\bf 11} 3529
\item[] Berakdar J, Briggs J S and Klar H 1989 \jpb {\bf 26} 285
\item[] Ciappina M F, Cravero W R and Garibotti C R 2003 \jpb {\bf 36} 3775
\item[] Ciappina M F, Cravero W R and Garibotti C R 2004 \PR A {\bf 70} 062713
\item[] Clemente E and Roetti C 1974 {\it At. Data Nucl. Data Tables} {\bf 14} 177
\item[] Crothers D S F 1982 \JPB {\bf 15} 2061
\item[] Crothers D S F and McCann J F 1983 \JPB {\bf 16} 3229
\item[] Crothers D S F and Dub\'e L J 1992 {\it Adv. At. Mol. Opt. Phys.} {\bf 30} 287-337
\item[] Dub\'e L J and Dewangan D P 1995 {\it 19th Int. Conf. on Physics of Electronic and Atomic Collisions (Whistler)} Abstracts p 62
\item[] Fainstein P D, Ponce V H and Rivarola R D 1991 \jpb {\bf 24} 3091
\item[] Fainstein P D and Guly\'{a}s L \jpb {\bf 38} (2005) 317
\item[] Fischer D, Moshammer R, Schulz M, Voitkiv A and Ullrich J 2003 \jpb {\bf 36} 3555
\item[] Foster M, Madison D H, Peacher J L, Schulz M, Jones S, Fischer D, Moshammer R and Ullrich J 2004 \jpb {\bf 37} 1565
\item[] Garibotti C R and Miraglia J E 1980 \PR A {\bf 21} 572
\item[] Guly\'as L, Fainstein P D and Salin A 1995 \jpb {\bf 28} 245
\item[] Guly\'as L and Fainstein P D 1998 \jpb {\bf 31} 3297
\item[] Inokuti M 1971 {\it Rev. Mod. Phys.} {\bf 43} 297
\item[] McCarroll R and Salin A 1978 \jpb {\bf 11} L693
\item[] Moshammer R, Ullrich J, Unverzagt M, Schmitt W, Jardin P, Olson R E, Mann R, D\"orner R, Mergel V, Buck U and Schmidt-B\"ocking H 1994 \PRL {\bf 73} 3371
\item[] Pedersen J O, Hvelplund P, Petersen A G and Fainstein P D 1990 \jpb {\bf 23} L597
\item[] Rosenberg L 1973 {\it Phys. Rev.} D {\bf 8} 1833
\item[] Schulz M, Moshammer R, Madison D H, Olson R E, Marchalant P, Whelan C T, Walters H R J, Jones S, Foster M,
Kollmus K, Cassimi A and Ullrich J 2001 \jpb {\bf 34} L305
\item[] Stolterfoht N, DuBois R D and Rivarola R D 1997 {\it Electron Emission in Heavy Ion-Atom Collisions} (Springer:Berlin)
\end{harvard}

\Figures

\begin{figure}
\caption{\label{figure1} FDCS for 100 MeV amu$^{-1}$ C$^{6+}$ single
ionization of Helium calculated in prior CDW--EIS: solid line;
experimental data, (Schulz {\it et al} 2001) solid circles. (a) $E_{e}=6.5$ eV, $|q|=0.88$ a.u. (b)
$E_{e}=17.5$ eV, $|q|=1.43$ a.u. (c) $E_{e}=37.5$ eV, $|q|=2.65$ a.u. }
\end{figure}

\begin{figure}
\caption{\label{figure2} FDCS for 100 MeV amu$^{-1}$ C$^{6+}$ single
ionization of Helium calculated for $E_{e}=17.5$ eV and $|q|=1.43$ a.u.
in prior CDW--EIS: solid line; post CDW--EIS: dashed line; FBA: dotted line; experimental
data: (Schulz {\it et al} 2001) solid circles. }
\end{figure}

\begin{figure}
\caption{\label{figure3} FDCS for 3.6 MeV amu$^{-1}$ Au$^{24+}$ single
ionization of Helium for $E_{e}=4$ eV.
Prior CDW--EIS: solid line; post CDW--EIS: dashed
line; experimental data: (Fischer {\it et al} 2003) solid circles. Note that
the angle of electron emission has been changed with respect to the other figures and
now is measured in a range of $-180^{\circ}$ to
$+180^{\circ}$, being $0^{\circ}$ the direction of the incoming projectile. }
\end{figure}

\begin{figure}
\caption{\label{figure4} Same as in figure 3 for $E_e=10$ eV.}
\end{figure}

\begin{figure}
\caption{\label{figure5} FDCS for 3.6 MeV amu$^{-1}$ Au$^{53+}$ single
ionization of Helium for $E_{e}$=4 eV.
Prior CDW--EIS: solid line; post CDW--EIS: dashed
line; experimental data: (Fischer {\it et al} 2003) solid circles. }
\end{figure}

\begin{figure}
\caption{\label{figure6} Same as in figure 5 for $E_{e}$=10 eV.}
\end{figure}

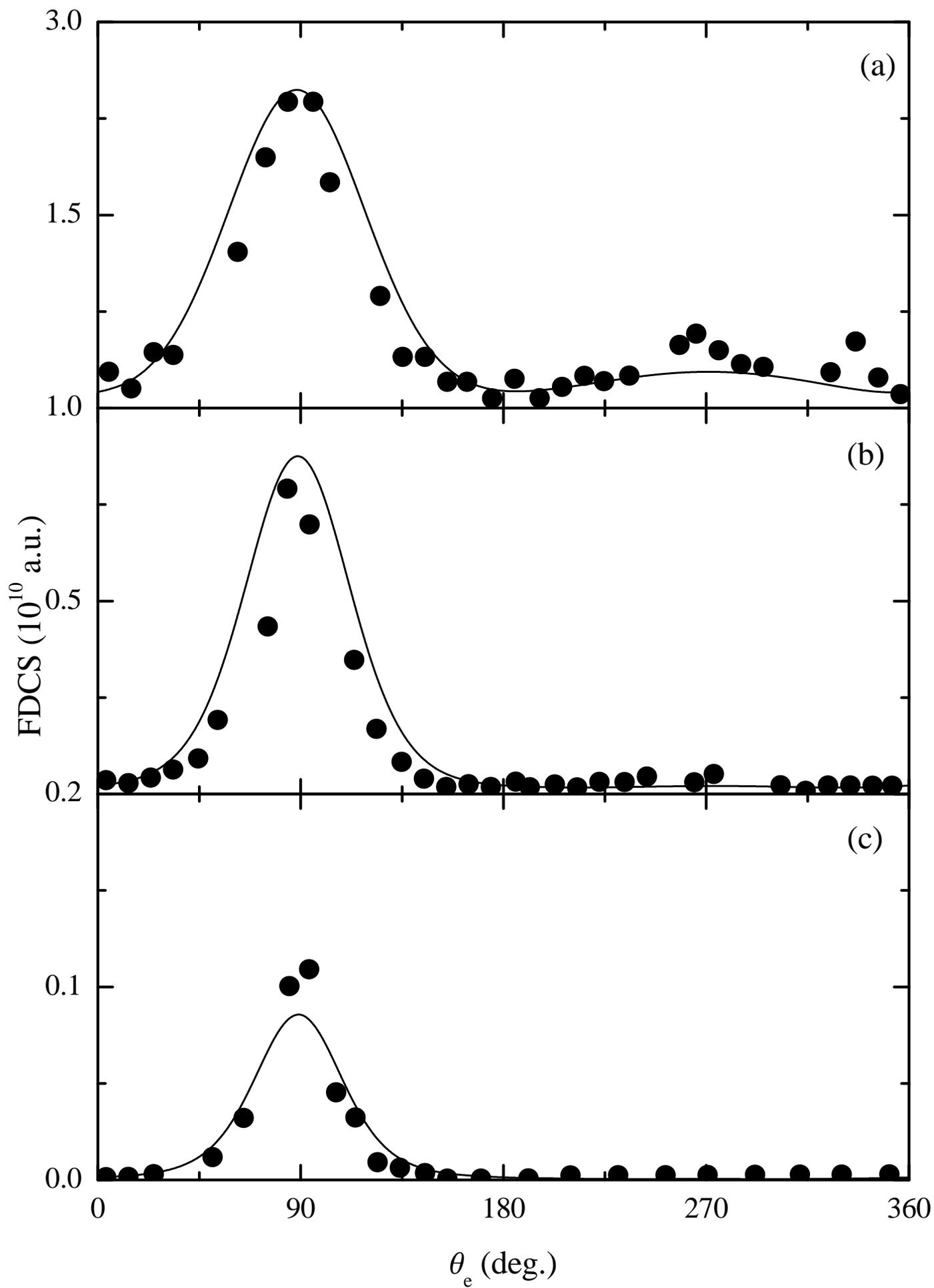
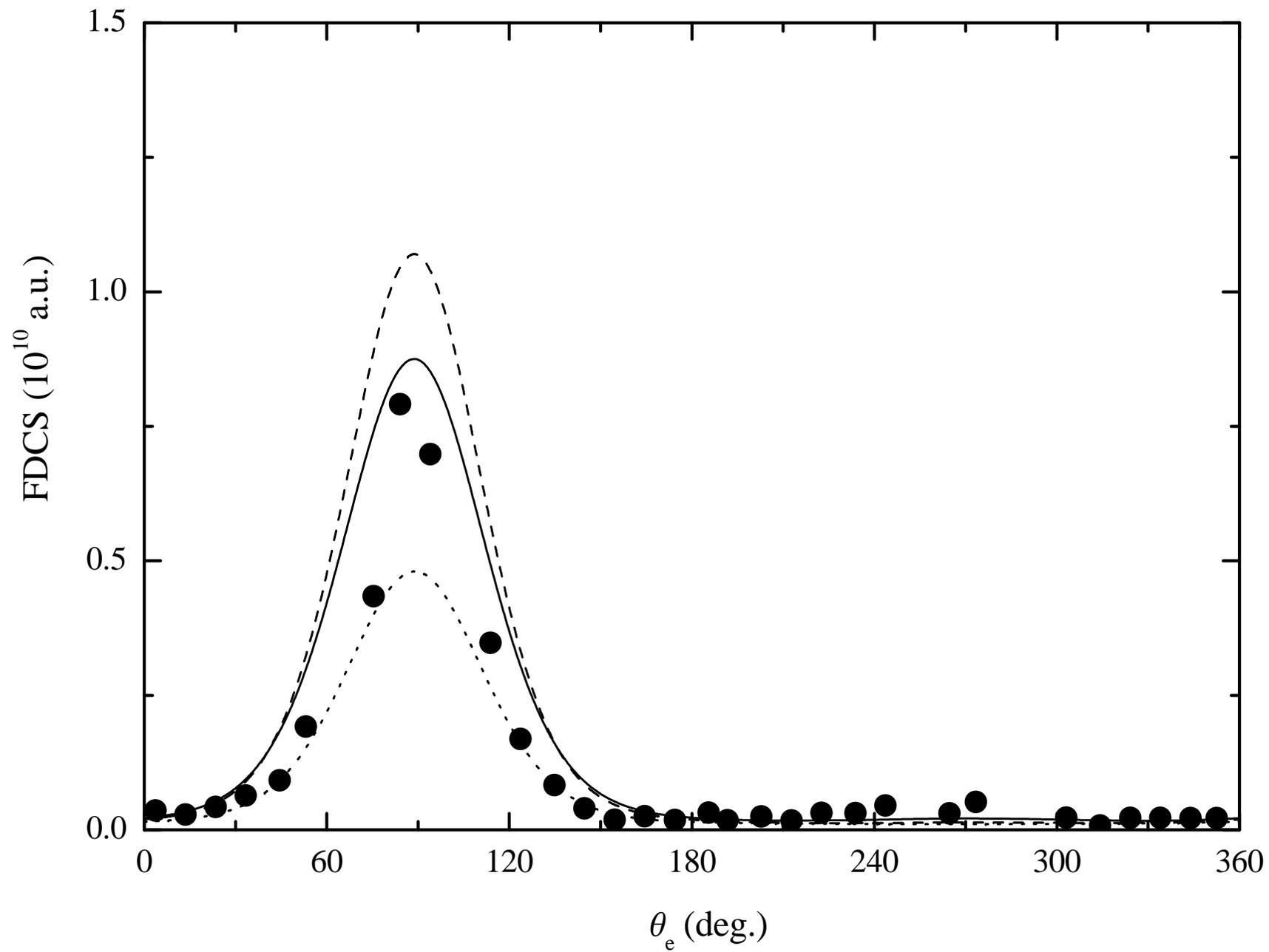
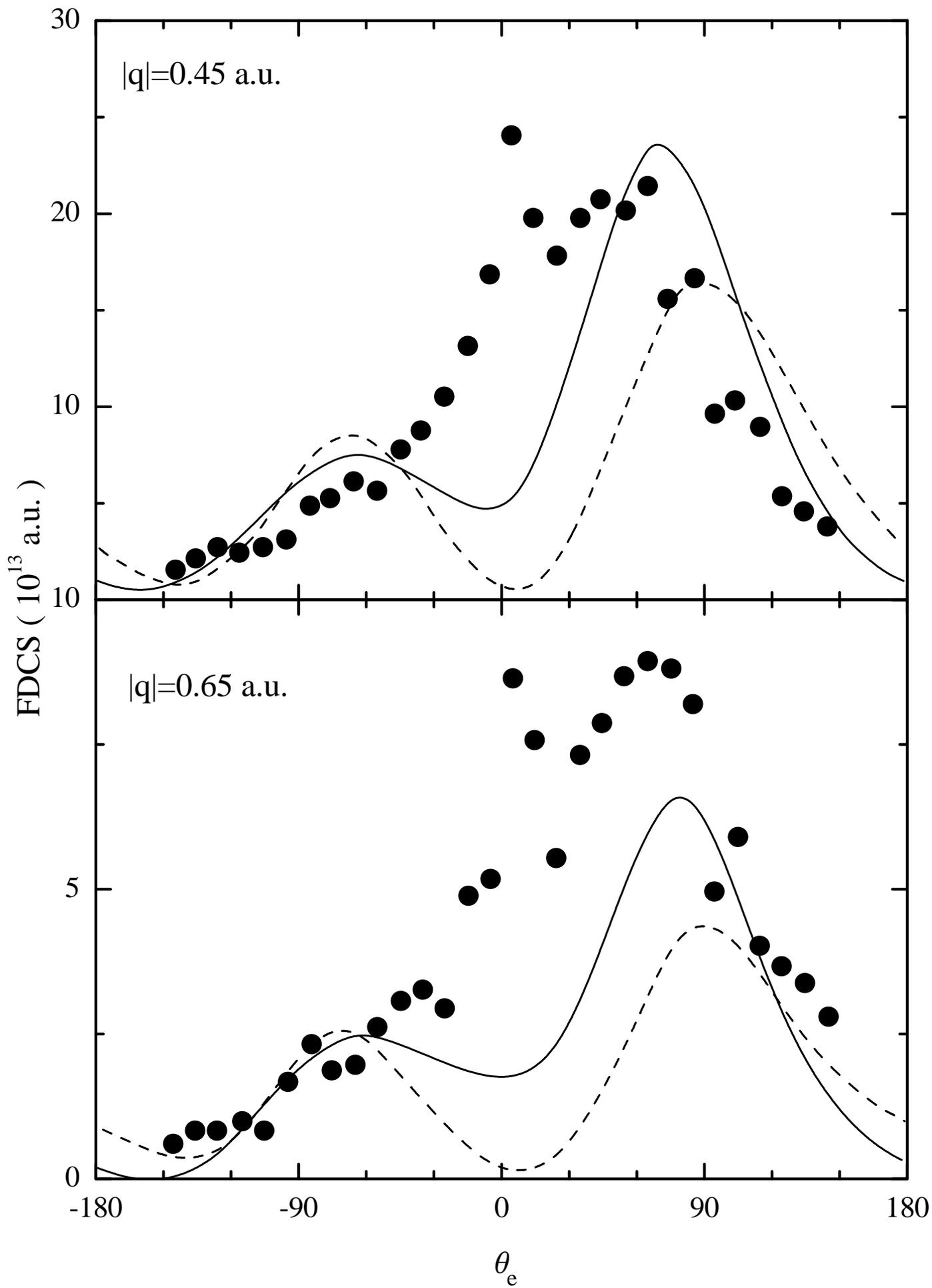
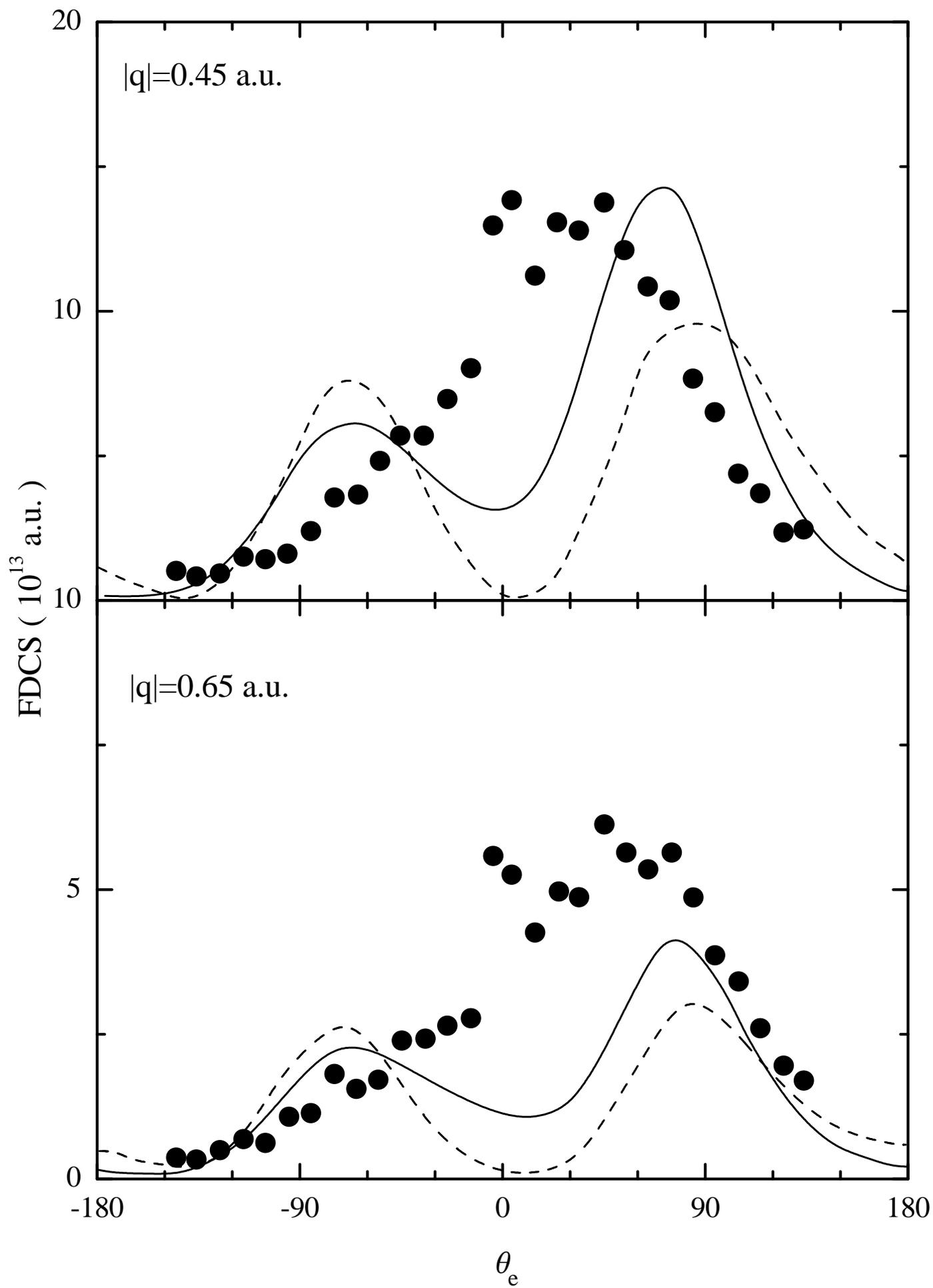
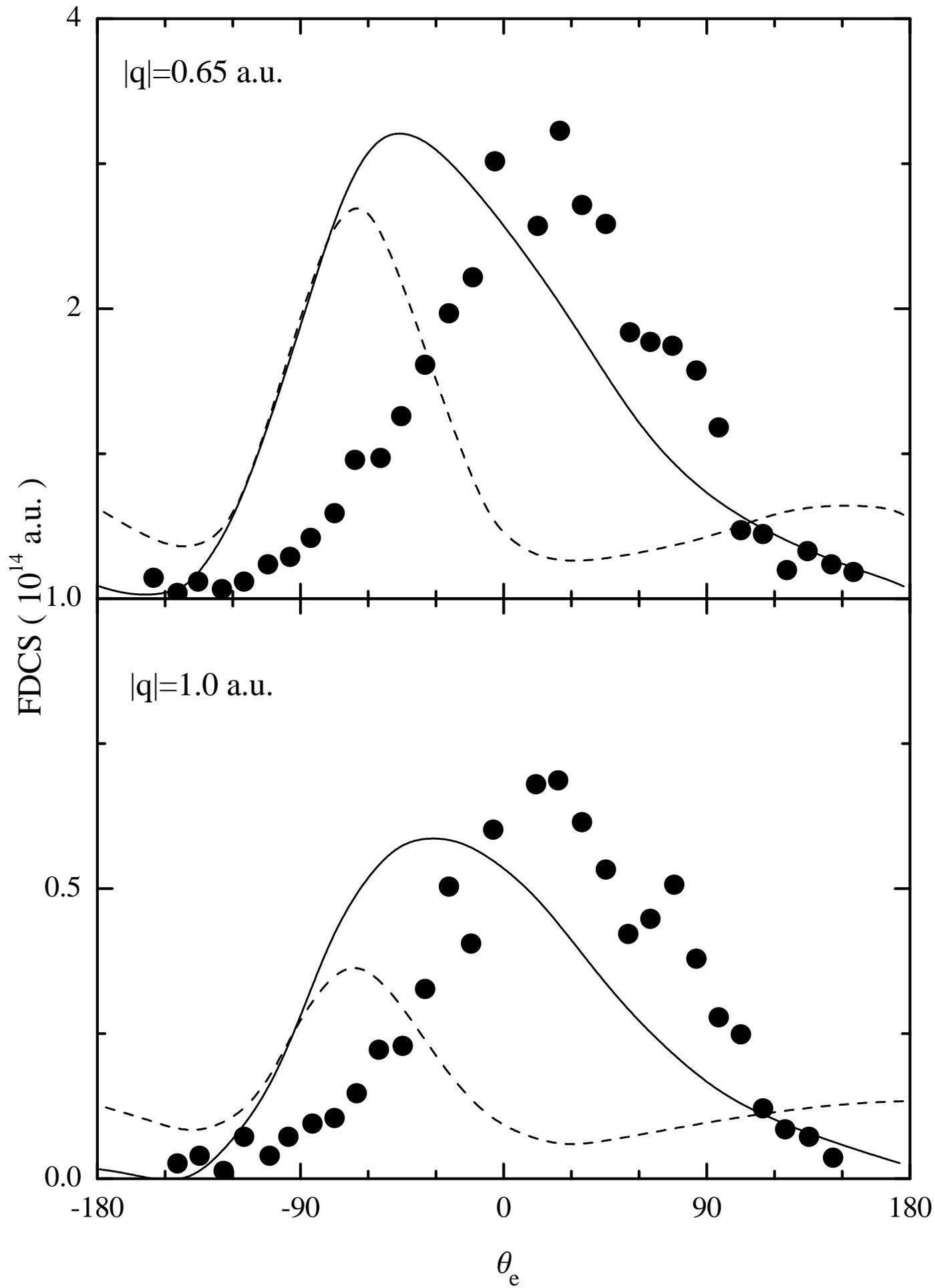
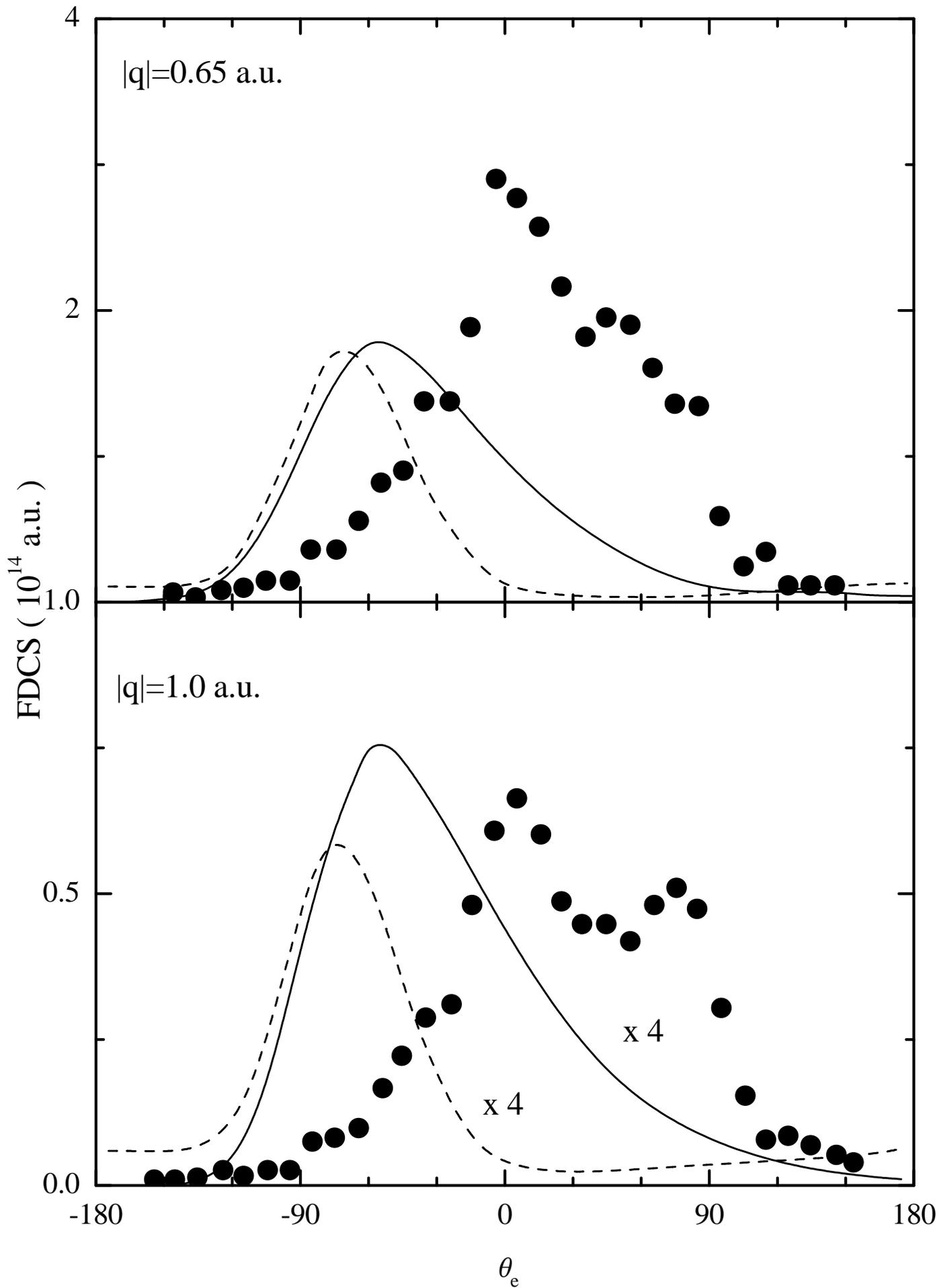
\end{document}